\documentclass[floatfix,twocolumn,aps,prd,showpacs,amsmath,amssymb]{revtex4}
\usepackage{bm}
\usepackage{epsfig}
\newcommand{\be}{\begin{equation}}
\newcommand{\ee}{\end{equation}}
\begin{document}


\title{Relativistic Arquimedes law for fast moving bodies and the \\
       general-relativistic resolution of the ``submarine paradox"}

\author{George E.\ A.\ Matsas}
\email{matsas@ift.unesp.br} 
\affiliation{Instituto de F\'\i sica Te\'orica, 
             Universidade Estadual Paulista,
             Rua Pamplona 145, 01405-900, 
             S\~ao Paulo, SP, Brazil}

\date{\today}

\begin{abstract}
We investigate and solve in the context of General Relativity the
apparent paradox which appears when bodies floating in a background 
fluid are set in relativistic motion. Suppose some macroscopic body, 
say, a submarine designed to lie just in equilibrium when it rests 
(totally) immersed in a certain background fluid.  The puzzle arises 
when different observers are asked to describe what is expected to 
happen when the submarine is given some high velocity parallel to the 
direction of the fluid surface. On the one hand, according to observers at 
rest with the fluid, the submarine would contract and, thus, sink as 
a consequence of the density increase. On the other hand, mariners at 
rest with the submarine using an analogous reasoning  for the 
fluid elements would reach the opposite conclusion. The general 
relativistic extension of the Arquimedes law for moving bodies shows 
that the submarine sinks. 
\end{abstract}

\maketitle

Suppose a submarine designed to lie just in equilibrium when it rests 
(totally) immersed in a certain background fluid. The puzzle appears 
when different observers are asked to describe what is expected to 
happen when the submarine is given some high velocity parallel to the 
direction of the fluid surface. On the one hand, according to observers at 
rest with the fluid, the submarine would contract and sink as a consequence 
of the density increase. On the other hand, mariners at rest with the 
submarine using an analogous  reasoning for the fluid elements would 
reach the opposite conclusion. To the best of our knowledge, the first
one to discuss this apparent paradox was Supplee~\cite{S}. Because
his analysis was performed in the context of Special Relativity,  
assumptions about how the Newtonian gravitational field would transform in 
different reference frames were unavoidable. In order to set the resolution 
of this puzzle on more solid bases, a general-relativistic 
analysis is required. We will adopt hereafter natural units:
$c=\hbar=G=k=1$, and spacetime metric signature $(-,+,+,+)$.

Let us begin writing the line element of the most 
general spherically symmetric static spacetime as 
\begin{equation}
ds^2 = -f(r) dt^2 + g(r) dr^2 +r^2 (d\theta^2 + \sin \theta^2  d\phi^2) \;,
\label{sss}
\end{equation} 
where $f(r)$ and $g(r)$ are determined by the Einstein equations
$G_{\mu \nu} = 8 \pi T_{\mu \nu}$. We will consider the base planet
where the experiment will take place as composed of two layers: an 
interior solid core with total mass $M$ and $r \in [0,R_-]$ ($R_- > 2M$) 
and an exterior liquid shell  with $ r \in (R_-, R_+]$. 
The gravitational field on the liquid shell will be assumed to be mostly 
ruled by the solid core, as verified, e.g., on Earth. In this case, the 
proper acceleration experienced by the  static liquid volume elements 
can be approximated by  $(M/r^2)/ \sqrt{1-2M/r}$ and, thus, 
{\em increases with depth}. 
This physical feature will be kept as we model the gravitational field 
on the fluid in which the submarine is immersed, but rather than locating 
it in the spacetime described by  Eq.~(\ref{sss}), we will look for a 
background with planar symmetry. This is necessary in order to avoid 
the appearance of centrifugal effects which are not part of the submarine 
paradox. This is accomplished by the Rindler spacetime 
\begin{equation}
ds^2 = e^{2 \alpha  {Z}} (-d{T}^2 + d{Z}^2) + dx^2 + dy^2 \;,
\label{rw}
\end{equation}
where $\alpha = {\rm const} > 0$. 
The liquid layer will be set at ${Z} \in (Z_-, 0]$, where $Z_- < 0$ and we
will assume that $|Z_-| \gg 1/\alpha$ in which case the total proper depth 
as defined by static observers will be approximately $1/\alpha$. The 
proper acceleration of the liquid volume elements at some
point $({T}, {Z}, x, y)$ is $a_{\rm (l)} = \alpha e^{-\alpha{Z}}$
and, thus, indeed increases as one moves to the bottom.

Let us assume the submarine to have rectangular shape and 
to lie initially at rest in the region $x>0$ at 
$[{Z}_\bot, {Z}_\top] \times [x_\vdash , x_\dashv] \times [y_1 , y_2]$. 
For the sake of simplicity, we will  assume the submarine to 
be thin with respect to the depth $1/\alpha$,
i.e. $e^{\alpha Z_\top} - e^{\alpha Z_\bot} \ll 1$. 
This is not only physically desirable as a way to minimize turbulence 
and shear effects, but also technically convenient as will be seen 
further. At ${T} = 0$ it begins to move along 
the $x$-axis towards increasing $x$ values in such a way that eventually 
its points acquire uniform motion characterized by the 3-velocity 
$v_0 \equiv dx/dT= {\rm const} > 0$. However, in order to keep 
the submarine uncorrupted, the whole process must be conducted 
with caution. First of all, we will impose that the 4-velocity 
$u_{\rm (s)}^\mu$ of the submarine points satisfy the {\em no-expansion 
condition}: $ \Theta\equiv \nabla_\mu u_{\rm (s)}^\mu = 0$. 
This can be implemented by the following choice:
\begin{equation}
u_{\rm (s)}^\mu = \frac{\chi^\mu + v(x^\alpha) \zeta^\mu}
                {|\chi^\mu + v(x^\alpha) \zeta^\mu)|}  \;,
\label{us}
\end{equation}
where $\chi^\mu = (1,0,0,0) $ and $\zeta^\mu = (0,0,1,0)$ are timelike and 
spacelike Killing fields, respectively, and
\begin{equation}
v (x^\alpha) \equiv \frac{dx}{dT} = \left\{
\begin{array}{ll}
0 & {\rm for}\;\;\;\;\; {T}/x < 0 \\
e^{2\alpha{Z}} {T}/x & {\rm for} \;\;\;\;\; 
                   0 \leq {T}/x \leq v_0 e^{-2\alpha {Z}}\\
v_0 & {\rm for}\;\;\;\;\; {T} /x > v_0 e^{-2 \alpha {Z}}  
\end{array}
\right. \; .
\label{v}
\end{equation}
Hence, a generic submarine  point will have 
a timelike trajectory in the region $x>0$, given by $Z=Z_0 = {\rm const}$, 
$y=y_0 = {\rm const}$ and 
\begin{equation}
x (T) =\left\{
\begin{array}{cl}
x_0   &  {\rm for}\;\;\;\;\; {T}< 0 
\\
\sqrt{x_0^2+e^{2\alpha{Z}_0}{T}^2 \,}
&{\rm for}\;\;\;\;\; 0 \leq{T}\leq {T}_{\rm un} ,
\\
x_0\sqrt{1-v_0^2 e^{-2\alpha Z_0}} + v_0 T & 
{\rm for}\;\;\;\;\; {T} > {T}_{\rm un}
\end{array}
\right. 
\label{x}
\end{equation}
where 
$T_{\rm un} = x_0 v_0 e^{-2\alpha {Z_0}}/\sqrt{1-v_0^2 e^{-2\alpha {Z_0}}}$
defines the moment after which each submarine point acquires uniform 
motion with constant 3-velocity $v_0$ ($0 < v_0 < e^{\alpha {Z}_0} $).  

It should be noticed that the no-expansion requirement is a necessary
but not sufficient condition to guaranty
that the submarine satisfies the rigid body condition
\begin{equation} 
\nabla^{( \mu } u_{\rm (s)}^{ \nu )} 
 + a_{\rm (s)}^{ ( \mu } u_{\rm (s)}^{ \nu )} = 0 \;,
\label{corpo rigido}
\end{equation}   
i.e. that the {\em proper} distance among the submarine points 
are kept immutable, where 
$a_{\rm (s)}^\mu \equiv u_{\rm (s)}^{ \nu } \nabla_\nu u_{\rm (s)}^{ \mu } $. 
This can be seen by recasting Eq.~(\ref{corpo rigido}) 
in the form 
\begin{equation}
\sigma_{\mu \nu} + (\Theta/3) P_{\mu \nu} = 0, 
\label{corpo rigido 2}
\end{equation}
where 
$
P_{\mu \nu} \equiv g_{\mu \nu} + u^{\rm (s)}_{ \mu } u^{\rm (s)}_{ \nu }
$
is the projector operator and
$$
\sigma_{\mu \nu} \equiv 
\left( \nabla_\alpha u^{\rm (s)}_{ (\mu  } \right) P^\alpha_{\;\; \nu)} - 
(\Theta/3) P_{\mu \nu}
$$ 
is the shear tensor. 
If the submarine were infinitely thin ($Z_\bot = Z_\top$), then
$\sigma_{\mu \nu}$ would vanish in addition to $\Theta$ and the
rigid body equation~(\ref{corpo rigido 2}) would be precisely 
verified. But
this is not so because the fact that $Z_\bot \neq Z_\top$  induces shear
as the submarine is {\em in the transition region:} 
$0 \leq T \leq T_{\rm un}$. 

In order to figure out how this can be minimized, 
we must first calculate the eigenvalues $\lambda_{\rm (i)}$ (i=1,2,3) and the 
corresponding (mutually orthogonal) spacelike eigenvectors 
$w^\mu_{\rm (i)}$ (which also satisfies  $w^\mu_{\rm (i)} u_\mu^{\rm (s)}  =0$) 
associated with the equation
$\sigma^\mu_{\;\;\nu} w_{\rm (i)}^\nu = \lambda_{\rm (i)} w_{\rm (i)}^\mu $: 
$$
\lambda_{\rm (1)} = 0 , \;\;\;\;\;\;\;\;
\lambda_{{\rm (2)}/{\rm (3)}}  = +/- \sqrt{ {\sigma^2} \,} ,
$$  
$$ 
w_{\rm (1)}^\mu = (0,0,0,1) , \;\;\;  
w_{{\rm (2)}/{\rm (3)}}^\mu = (\sigma^0_{\;\;1}, 
                  +/- \sqrt{\sigma^2\, } ,
                  \sigma^2_{\;\; 1},
                  0) \; ,
$$  
where 
$$
\sigma^2 \equiv 
\sigma^{\mu \nu} \sigma_{\mu \nu}/2 = 
a_{\rm (l)}^2 x^2 (x^2 - x_0^2)/x_0^4 \;,
$$
$$
\sigma^0_{\;\;1} =  \alpha e^{-\alpha Z} x(x^2 - x_0^2)/x_0^3, \;
\sigma^2_{\;\;1} = \alpha x^2 (x^2 - x_0^2)^{1/2}/x_0^3  
$$
and
we recall that $a_{\rm (l)} = \alpha e^{-\alpha Z}$.
Then, by locally choosing  a 3-vector basis 
$e_{\rm (i)}^\mu = w_{\rm (i)}^\mu$ and 
assuming that $e_{\rm (i)}^\mu$ is orthogonally transported along 
$u_{\rm (s)}^{ \mu }$, i.e. 
$
[u_{\rm (s)}, e_{\rm (i)}]^\mu = 
a^{\rm (s)}_\nu e_{\rm (i)}^\nu u_{\rm (s)}^\mu 
$, 
one obtains
$
u_{\rm (s)}^\mu \nabla_\mu |e_{\rm (i)}^\nu| = 
\lambda_{\rm (i)} |e_{\rm (i)}^\nu|  
$.
Hence, the distortion rate  
of a sphere inside the submarine along the principal axes 
$e_{\rm (i)}^\mu$ is given by the corresponding eigenvalues 
$\lambda_{\rm (i)}$. In our case, no distortion is verified along 
the $y$-axis (see $\lambda_{\rm (1)}$ and 
$w_{\rm (1)}^\mu$) and the distortion which appears in the 
transition region associated with the $Z$-axis
can be minimized by making 
$|\lambda_{{\rm (2)}/{\rm (3)}}|$ small 
enough. By using Eq.~(\ref{x}) (at $T = T_{\rm un} $), one obtains  
$$
 |\lambda_{\rm (2)}| = |\lambda_{\rm (3)}| \leq 
 a_{\rm (l)} v_0 e^{- \alpha Z_\bot}/(1-v_0^2 e^{-2\alpha Z_\bot}) \;.
$$
Thus one can minimize shear effects in the submarine either
(i) by making the final velocity to be moderate ($v_0 \ll 
e^{\alpha Z_\bot}$), (ii) by setting it in a small-acceleration 
region [in comparison to the inverse of the submarine 
$Z$-proper size: 
$a_{\rm (l)} \ll \alpha/(e^{\alpha Z_\top} - e^{\alpha Z_\bot} )]$, 
or, as considered here, (iii) by designing the submarine thin 
enough ($e^{\alpha Z_\top} - e^{\alpha Z_\bot} \ll 1$).

\begin{figure}
\epsfig{file=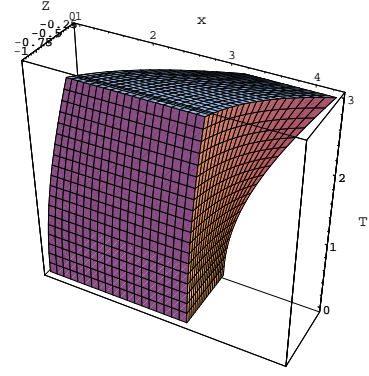,angle=0,width=0.55 \linewidth,clip=}
\caption{\label{gg} The time evolution of a $y = {\rm const}$ 
section is plotted (using $\alpha =1$). At $T=0$, the submarine is at 
rest and all the observers agree about its rectangular 
shape. As its velocity increases, however, the submarine contracts as
a function of $Z$ according to the observers at rest with the fluid 
[more on the top than at the bottom (see slices $T= {\rm const} >0$)],
although mariners aboard in will detect no relevant change in shape.
}
\end{figure}

After the transition region, all the submarine points will follow 
isometry curves associated with the timelike Killing field 
$\eta^\mu = \chi^\mu + v_0 \zeta^\mu$. It is easy to check by using 
\begin{equation}
a_{\rm (s)}^\mu =(\nabla^\mu \eta)/ \eta =
(0,\alpha e^{-2\alpha {Z}}/(1-v_0^2 e^{-2 \alpha{Z}}),0,0) \; , 
\label{4-acceleration}
\end{equation}
where 
$\eta \equiv |\eta^\mu| 
       = e^{ \alpha {Z}} (1 - v_0^2 e^{- 2 \alpha {Z}} )^{1/2}
$ 
that the rigid body equation is fully verified in this stationary 
region: $T>T_{\rm un}$. 
It is interesting to notice that although mariners aboard
will not perceive any significant change in the submarine's form, 
observers at rest with the fluid will witness a relevant contraction
in the $x$-axis direction as a function of 
${Z}$ (and $v_0$); indeed, more at the top than at the bottom 
(see Fig.~\ref{gg}). 

Now, let us suppose that the liquid layer in which the submarine is 
immersed is a perfect fluid characterized by the energy-momentum 
tensor 
$$
T^{\mu \nu} = 
              \rho_{\rm (l)} u_{\rm (l)}^\mu u_{\rm (l)}^\nu + 
              P_{\rm (l)} (g^{\mu \nu} + u_{\rm (l)}^\mu u_{\rm (l)}^\nu) \;,
$$ 
where 
$
u_{\rm (l)}^\mu = \chi^\mu / \chi 
\label{ul}
$
with 
$ \chi = |\chi^\mu | = e^{\alpha {Z}}$,
and 
$\rho_{\rm (l)}$ and  $P_{\rm (l)}$ are the fluid's proper energy 
density and pressure, 
respectively. From $\nabla_\mu T^{\mu \nu} = 0$, we obtain 
\begin{equation}
\nabla^\mu P_{\rm (l)} = -(\rho_{\rm (l)} + P_{\rm (l)}) a_{\rm (l)}^\mu \;, 
\label{eqdofluido}
\end{equation}
where 
$
a_{\rm (l)}^\mu = (0, \alpha e^{-2\alpha {Z}},0,0) 
$.
For later convenience, we cast Eq.~(\ref{eqdofluido}) in the form
\begin{equation}
\rho_{\rm (l) }  {d\chi}/{ dl} + { d(\chi  P_{\rm{(l)}})}/{dl} =0 \;, 
\label{eqdofluido2} 
\end{equation}
where we have used that 
$
a^\mu_{\rm (l)} = (\nabla^\mu \chi) / \chi
$
and
$dl$ is the differential proper distance in the ${Z}$-axis direction. 

The {\em proper} hydrostatic pressures at the bottom $P_\bot$ and 
on the top $P_\top$ of the submarine will be given by 
\begin{equation}
P_{\bot/\top} \equiv T_{\mu \nu} N_{\bot/\top}^\mu N_{\bot/\top}^\nu 
                 = P_{\rm (l)}|_{{Z} = {Z}_{\bot/\top}} \; ,
\end{equation}
where 
$N_{\bot/\top}^\mu = (0,1,0,0) e^{-\alpha {Z}_{\bot/\top}}$ 
are unit vectors orthogonal to the submarine's 4-velocity
(and to the top and bottom surfaces). Thus, the hydrostatic scalar 
forces on the top and at the bottom of the submarine are   
$$
F_{\bot/\top} = +/\!- A P_{\bot/\top} 
              = +/\!- A P_{\rm (l)}|_{{Z} = {Z}_{\bot/\top} }\; ,
$$ 
where $A$ is  the corresponding proper area. 

In order to combine $F_\bot$ 
and $F_\top$ properly, we must transmit them to a common holding point.
Let us assume that the forces are transmitted through a lattice of ideal 
cables and rods to some arbitrary inner point  
${\cal O} \equiv (Z_{\cal O}  ,x_{\cal O}  ,y_{\cal O})$ inside the 
submarine, where
its mass is also concentrated. Ideal cables and rods are those 
ones which transmit pressure through $\nabla_\mu T^{\mu \nu} = 0$ and 
have negligible energy densities. As a consequence of our thin-submarine 
assumption, our final answer will be mostly insensitive to the 
choice of ${\cal O}$.  $F_{\bot/\top}$ are related to
the transmitted forces $F_{\bot/\top}^{\cal O}$ at ${\cal O}$ by 
$
F_{\bot/\top}^{\cal O} = 
[\eta ({Z}_{\bot/\top})/\eta({Z}_{\cal O})] F_{\bot/\top} \;.
$
Hence, the Arquimedes law induces the following scalar force 
(along the ${Z}$-axis) at 
${\cal O}$ \begin{equation}
F^{\cal O}_{\rm A} 
  =  F_{\bot}^{\cal O} + F_{\top}^{\cal O} 
  = - \left. \frac{V}{\eta ({Z} )} 
       \frac{d(\eta ({Z}) P_{\rm (l)})}{dl}
        \right|_{{Z} = {Z}_{\cal O}}  \;,
\label{Fa}
\end{equation} 
where $V$ is the submarine's proper volume and we have assumed that  
$d(\eta ({Z}) P_{\rm (l)})/dl$ does not vary much along the submarine so 
that we can neglect higher derivatives. This is natural in light
of our thin-submarine assumption.

In addition to $F^{\cal O}_{\rm A}$, we must consider the force 
(along the ${Z}$-axis) 
associated with the gravitational field:
\begin{eqnarray}
F^{\cal O}_{\rm g} 
& = & - m a_{\rm (s)}^{\mu} N_\mu|_{{Z} = {Z}_{\cal O}}
\nonumber \\
& = & 
- m N^\mu (\nabla_\mu \eta) /\eta |_{{Z} = {Z}_{\cal O}}
\nonumber \\
& = & 
- \left. ({m}/{\eta ({Z})}) ({ d \eta ({Z})}/{dl}) 
  \right|_{{Z} = {Z}_{\cal O}}\;, 
\label{Fg}
\end{eqnarray} 
where 
$ a_{\rm (s)}^{\mu}|_{{Z} = {Z}_{\cal O}} $
is obtained from Eq.~(\ref{4-acceleration}), 
$m$ is the submarine mass and
$N_\mu|_{{Z} = {Z}_{\cal O}}= (0,1,0,0) e^{\alpha {Z}_{\cal O}}$.

Now, by adding up Eqs.~(\ref{Fa}) and (\ref{Fg}) we obtain the total force
on the submarine as
\begin{equation}
F^{\cal O}_{\rm tot} = 
- \left[ \frac{m}{\eta ({Z})} \frac{d\eta ({Z})}{dl} 
+ \frac{V}{\eta ({Z})} \frac {d(\eta ({Z}) P_{\rm{(l)}})}{dl}
  \right]_{{Z} = {Z}_{\cal O}} .
\label{Ftot}
\end{equation} 
In order to fix the submarine's mass, we give to it just the necessary 
ballast to keep it in hydrostatic equilibrium when it lies 
at rest completely immersed. This means that we must impose
$
F^{\cal O}_{\rm tot}|_{v_0=0} = 0.
\label{balance}
$
Now, by recalling that $\eta \stackrel{v_0 \to 0}{\longrightarrow} \chi$
and using Eq.~(\ref{eqdofluido2}), we reach the 
conclusion~\cite{UW} that the equilibrium condition above
implies that the submarine must be designed such that its 
mass-to-volume ratio obey the simple relation
$ m/V = \rho_{\rm (l)}$.
Then, by using this and Eq.~(\ref{eqdofluido2}) in Eq.~(\ref{Ftot}), 
it is not difficult to write the total proper force on the moving submarine 
as 
\begin{eqnarray}
F^{\cal O}_{\rm tot} 
& = & \left.
   - V ( \rho_{\rm (l)} + P_{\rm (l)} ) 
    \left(
    \frac{1}{\eta} \frac{d\eta}{dl} - \frac{1}{\chi} \frac{d\chi}{dl} 
    \right)
      \right|_{{Z} = {Z}_{\cal O}}
\nonumber \\
& = & \left.
    - V ( \rho_{\rm (l)} + P_{\rm (l)} ) N^\mu 
    \left(
    \frac{\nabla_\mu \eta}{\eta} - \frac{\nabla_\mu \chi}{\chi } 
    \right)
    \right|_{{Z} = {Z}_{\cal O}}
\nonumber 
\end{eqnarray}
and, thus,
\begin{equation}
F^{\cal O}_{\rm tot} = 
    \left.
    \frac{- V ( \rho_{\rm (l)} + P_{\rm (l)} ) 
    a_{\rm (l)} v_0^2 e^{-2 \alpha {Z}}}{1- v_0^2 e^{-2 \alpha {Z}} }
    \right|_{{Z} = {Z}_{\cal O}} \; ,
\label{Ftotv}
\end{equation} 
where we recall that $a_{\rm (l)} = \alpha e^{-\alpha Z}$.
Clearly, for $v_0 =0$ we have $F^{\cal O}_{\rm tot} = 0$, as it 
should be, but for $v_0 \neq 0$ we have  
$F^{\cal O}_{\rm tot} < 0 $ and, thus, {\em we conclude that a net 
force downwards is exerted on the submarine}. 

In order to make contact of this result with the one obtained through
Special Relativity, let us begin by assuming
$\rho_{\rm (l)} = \rho_0 = {\rm const}$, in which case we can 
easily solve Eq.~(\ref{eqdofluido}):
$
P_{\rm (l)} = \rho_0 (e^{-\alpha {Z}} - 1)\; .
$
By letting this in Eq.~(\ref{Ftotv}), we obtain
\begin{equation}
F^{\cal O}_{\rm tot} = \left.
          - \frac{m \alpha v_0^2 e^{-4\alpha {Z}}}
          {1- v_0^2 e^{-2 \alpha {Z}} }
                       \right|_{{Z} = {Z}_{\cal O}} \; .
\label{Ftotvpart}
\end{equation} 
Now, let us assume that the submarine is close to the surface, i.e. 
at ${Z} \approx 0$, in which case
the line element (\ref{rw}) reduces to the usual line element form 
of the Minkowski space with $({T}, {Z}, x, y)$ playing the role of the 
Cartesian coordinates. As a consequence, 
Eq.~(\ref{Ftotvpart}) reduces to 
 \begin{equation}
F^{\cal O}_{\rm tot} \approx \left.
          - m g \gamma (\gamma - 1/\gamma)
                       \right|_{{Z} \approx 0} \; ,
\label{Ftotvpart2}
\end{equation}  
where $\gamma \equiv 1/ \sqrt{ 1-v_0^2 }$ and we have assumed that the
gravitational field is small enough such that we can identify the
proper acceleration 
on the liquid volume elements
$a_{\rm (l)} = \alpha e^{-\alpha {Z}} 
               \stackrel{{Z} \to 0}{\longrightarrow} \alpha$ 
with the {\em Newtonian} gravity acceleration $g$.
Notice that the first and second terms in Eq.~(\ref{Ftotvpart2}) 
can be associated with the proper gravitational 
and buoyancy forces, respectively.
Finally, by evoking Special Relativity to transform
the force from the proper frame of the submarine~(\ref{Ftotvpart2}) 
to the one at rest with the fluid, we reobtain
Supplee's formula~\cite{S}: 
$$
F_{\rm tot} = - m g (\gamma - 1/\gamma) \; .
$$    
Thus according to observers at rest with the fluid, the 
gravitational field on the 
moving submarine  increases effectively by a $\gamma$ factor 
as a consequence of the blue-shift on the submarine's energy 
and the buoyancy force  decreases by the same factor because 
of the volume contraction.
The apparently contradictory conclusion reached in the submarine rest frame 
by the mariners, who would witness a density increase of the liquid volume
elements is resolved by recalling that the 
gravitational field is not going to ``appear'' the same to them
as to the observers at rest with the fluid. This is naturally taken
into account in the General-Relativistic approach (and turned out
to be the missing ingredient which raised the paradox). This can be 
seen from Eq.~(\ref{Fg}) by casting it in the form
$
F^{\cal O}_{\rm g} = 
- m \alpha e^{- \alpha Z_{\cal O}}/(1-v_0^2 e^{-2\alpha Z_{\cal O}}) 
$.
Hence, according to mariners aboard, the effective
gravitational force will be larger when the submarine is 
moving than when it is at rest by a factor 
$(1-v_0^2 e^{-2\alpha Z_{\cal O}})^{-1} > 1$, pushing it
downwards. 

The Theory of Relativity is close to commemorate its first 
centennial. This is quite remarkable that it has not lost the
gift of surprising us so far. This is definitely a privilege 
of few elders. 

\begin{acknowledgments}

The author is thankful to J. Casti\~neiras, I. P. Costa e Silva and
D. A. T. Vanzella for general discussions.  
G.M. also acknowledges partial support from Conselho Nacional de 
Desenvolvimento Cient\'\i fico e Tecnol\'ogico and Funda\c c\~ao de 
Amparo \`a Pesquisa do Estado de S\~ao Paulo.

\end{acknowledgments} 

\end{document}